\documentclass[11pt, a4paper, twocolumn, copyright, goog]{google}

\usepackage{xspace}
\usepackage{arydshln}

\usepackage[authoryear, sort&compress, round]{natbib}
\uselogo{} 

\title{VaultGemma: A Differentially Private Gemma Model}

\correspondingauthor{amersinha@google.com}

\renewcommand{\today}{2025-09-12}
\newcommand{\dpgemma}{{VaultGemma 1B}\xspace}

\author[1, 2]{VaultGemma Team}

\affil[1]{Google Research}
\affil[2]{Google DeepMind}

\begin{abstract}
We introduce \dpgemma, a 1 billion parameter model within the Gemma family, fully trained with differential privacy. Pretrained on the identical data mixture used for the Gemma 2 series, \dpgemma\ represents a significant step forward in privacy-preserving large language models. We openly release this model to the community.
\end{abstract}

\begin{document}

\maketitle

\section{Introduction}
Large Language Models (LLMs) have demonstrated remarkable capabilities across a wide range of tasks, yet a significant challenge in their development and deployment is the inherent privacy risk. Trained on vast, web-scale corpora, LLMs have been shown to be susceptible to verbatim memorization and extraction of training data ~\citep{carlini2021extracting,carlini22quantifying,ippolito2022preventing,lukas2023analyzing,biderman2023emergent,prashanth2024recite}. This can lead to the inadvertent disclosure of sensitive or personally identifiable information (PII) that was present in the pretraining dataset.

To address these challenges, Differential Privacy (DP)~\citep{dwork2006calibrating} has emerged as the gold standard, providing a rigorous, mathematical framework to limit the influence of any single example in the training data on the resulting model. A model trained with DP provably bounds the reconstruction or leakage of information tied to individual data points.

An LLM encounters the vast majority of its training data during the initial pretraining phase~\citep{abdin2024phi,team2024gemma,team2024gemma2}. This stage relies on massive, heterogeneous datasets that, despite filtering efforts, can contain sensitive information. Thus, it is important to consider pretraining an LLM fully with DP. This approach provides an end-to-end privacy guarantee from the ground up, ensuring the foundational model is built in a way that prevents the memorization of specific, sensitive details. It allows the model to learn general patterns and knowledge about the world without being overly influenced by any single document or user's data, fundamentally mitigating the risk of privacy leaks from the original training corpus.

A common alternative to full private pretraining is to apply DP exclusively during the fine-tuning phase. However, this approach leaves the foundational model and its vast pretraining data unprotected, leading to several significant pitfalls. First, the model may have already memorized sensitive PII from the pretraining corpus before DP fine-tuning begins, and this process does not retroactively erase the memorized information. This leaves the model vulnerable to extraction attacks that can reveal verbatim training data—a risk that is significantly amplified for open-weight models, where adversaries have full access to model weights to probe for and reconstruct sensitive information. Consequently, this practice can create a false sense of security, as labeling the model ``private'' applies only to the fine-tuning data, while the core pretraining data remains at risk~\citep{positiontramer2022}.

VaultGemma represents a significant step forward in the journey toward building AI that is both powerful and private by design. By developing and applying a new, robust understanding of the scaling laws for DP~\citep{mckenna2025scaling}, we have successfully trained and released the largest open-weight, privately trained language model to date. Our primary motivation for releasing VaultGemma is to accelerate research and development in private AI. By providing the community with a powerful, high-utility private model and a clear methodology, we aim to lower the barrier to entry for building privacy-preserving technologies. Furthermore, this model can serve as a valuable foundation for applications where privacy of the training data when using the model is paramount.

While a utility gap still exists between private and non-private models, our work demonstrates that this gap can be systematically narrowed. We hope that VaultGemma and our accompanying research will empower the community to build the next generation of safe, responsible, and private AI for everyone.

\section{Model Architecture}
\label{sec:arch}
Similar to previous Gemma models~\citep{team2024gemma, team2024gemma2}, \dpgemma, is a  decoder-only transformer model, with most architecture elements similar to other Gemma versions.

\begin{table}[h]
\centering
\caption{Overview of the main parameters and design choices for the 1B model. See \autoref{sec:arch} for more details.}
\label{tab:model_parameters_2b}
\begin{tabular}{cc}
\hline
\textbf{Parameters} & \textbf{1B} \\
\hline
$d\_{\text{model}}$ & 1152 \\
Layers & 26 \\
Pre-norm & yes \\
Post-norm & no \\
\hline
Non-linearity & GeGLU \\
Feedforward dim & 13,824 \\
\hline
Head type & MQA \\
Num heads & 4 \\
Num KV heads & 1 \\
Head size & 256 \\
Global att. span & 1024 \\
\hline
Vocab size & 256,128 \\
Tied embedding & yes \\
\hline
\end{tabular}
\end{table}

\subsection{Sequence Length}
We choose to decrease the sequence length to 1024 for pretraining. We find that using a smaller sequence length significantly reduces compute requirements, which in turn allows us to train using larger batch sizes---a necessity for good performance in private training.

\subsection{Global Attention}
Given our use of a small sequence length, we choose to use global attention on all layers rather than alternating with sliding window attention. 

\subsection{Pre-norm with RMSNorm}
To stabilize training, we use RMSNorm~\citep{rms_norm} to normalize the input of each transformer sub-layer, the attention
layer, and the feedforward layer.

\section{Dataset}
We train using the same pretraining dataset as Gemma 2 27B. This dataset contains 13T  tokens of primarily-English data. These tokens come from a variety of data sources, including web documents, code, and science articles and only contain text data.~\citep{team2024gemma2}

\subsection{Filtering} We use the same data filtering techniques as ~\citep{team2024gemma2}. Specifically, we filter the pretraining dataset to reduce the risk of unwanted
or unsafe utterances, filter out certain personal information or other sensitive data, decontaminate evaluation sets from our pretraining data mixture, and reduce the risk of recitation by minimizing the proliferation of sensitive outputs.

\subsection{Tokenizer}
We use the same non-private tokenizer as Gemma 1, Gemma 2, and Gemini: a SentencePiece tokenizer~\citep{kudo2018sentencepiecesimplelanguageindependent} with split digits, preserved whitespace, and byte-level encodings. The resulting vocabulary has 256K entries. 

\section{Evaluations}
We show the performance of our final pretrained model across a variety of benchmarks and compare it against its non-private counterpart across a range of standard academic benchmarks in \autoref{tab:dp_model_comparison}. To put this performance in perspective and quantify the current impact of DP on performance, we also include a comparison to an older similar-sized GPT-2 model, which performs similarly on these benchmarks. This comparison illustrates that today’s private training methods produce models with utility comparable to that of non-private models from roughly five years ago, highlighting the important gap our work will help the community systematically close.

\begin{table*}[h!]
\centering
\caption{A comparison of DP and standard model performance and training configurations.}
\label{tab:dp_model_comparison}
\resizebox{\textwidth}{!}{%
\begin{tabular}{@{}lcccccccc@{}}
\toprule
\textbf{Model} & \textbf{ARC-C} & \textbf{ARC-E} & \textbf{HellaSwag} & \textbf{PIQA} & \textbf{SIQA} & \textbf{BoolQ} & \textbf{TriviaQA} \\
\multicolumn{1}{c}{} & 0 shot & 0 shot & 0 shot & 0 shot & 0 shot & 0 shot & 5 shot \\
\midrule
\dpgemma & 26.45 & 51.78 & 39.09 & 68.00 & 46.16 & 62.04 & 11.24 \\
\hdashline
Gemma3 1B (PT) & 38.31 & 71.34 & 61.04 & 77.37 & 49.28 & 68.75 & 39.75 \\
GPT-2-1.5B & 39.78 & 51.10 & 47.91 & 70.51 & - & 61.80 & 6.00 \\
\bottomrule
\end{tabular}%
}
\end{table*}

\section{DP Implementation}

\subsection{Private Training}

We implemented DP-SGD~\citep{DBLP:conf/ccs/AbadiCGMMT016} on top of the Gemma pretraining pipeline using clipping and noise addition components provided by JAX Privacy~\citep{jax-privacy2022github}. Our implementation uses vectorized per-example clipping for maximum parallelism, and gradient accumulation to simulate large batch sizes. Gradient accumulation steps are independent and each adds properly calibrated Gaussian noise to the partial gradients so that when these partial noisy gradients are averaged we obtain the target gradient required for DP-SGD model updates.

\subsection{Batch Construction}

\paragraph{Repeated Documents.}
Our training mixture is composed of a diverse set of documents sampled from numerous source datasets. We sample documents from these source documents into the mixture with probability proportional to its weight. As these datasets are of different quality, we assign different weights to sample documents from them into our mixture. At worst, we can sample a single document from these sources up to seven times in our mixture. However, for most of the source datasets, we sample documents fewer than three times into our final mixture.

\paragraph{Packing.}
To increase training efficiency, we pack documents into fixed size sequences of 1024 tokens. Due to the diversity of the type of documents we train on, our packing can pack multiple documents into a single example or break up a single document into multiple sequences. 

\paragraph{Privacy Guarantee.}
VaultGemma was trained with a ($\epsilon \le 2.0$, $\delta \le 1.1e^{-10}$)-sequence-level DP guarantee, where sequence consists of 1024 tokens after sampling and packing. It is important to note that if two repeated sequences occurred due to the sampling of repeated documents, they are treated as separate privacy units for the purpose of this guarantee.

\paragraph{Truncated Poisson Subsampling.}
We employ Truncated Poisson Subsampling~\citep{DBLP:conf/nips/ChuaGK0MSZ24} for sampling our mini-batches. This method provides a computationally efficient approximation of Poisson subsampling, where each example in the dataset is included in a batch with a fixed probability. Poisson subsampling results in batches of variable size which can result in slower training throughput. Truncated Poisson Subsampling allows us to use a fixed batch size by padding the batch when it is too small and truncating it when it is too large.

A key distinction in our methodology lies in the implementation of the data pipeline. The original work suggests a implementing truncated Poisson sub-sampling using  MapReduce. This approach is designed to minimize the overhead during training by front-loading the sampling computation. However, we found that a pre-generation step was not necessary for our use case and introduced complexities in data handling and storage.

Instead, we implement Truncated Poisson Sampling directly within our data loading pipeline using \texttt{pygrain}~\citep{grain2023github}. Our implementation performs on-the-fly sampling and batching of the data as it is fed to the model. We found that this approach does not introduce a significant computational overhead, and we observe data throughput speeds comparable to a standard data pipeline at the same physical batch size. We also find that overhead of padding is small at large batch sizes that we use for training with the padding accounting for less than 2\% of the total batch size.

\subsection{Privacy Accounting}

Our privacy accounting methodology is based on the $\text{ABLQ}_P$ method under the ``zeroing-out'' adjacency notion~\citep{pmlr-v139-kairouz21b} as detailed in \citep{DBLP:conf/nips/ChuaGK0MSZ24}. The accounting is implemented using the PLD accountant~\citep{DBLP:journals/popets/DoroshenkoGKKM22} from the Google DP accounting library \citep{dp_accounting}. 

\section{Scaling Law}

\subsection{Methodology}
Our methodology for deriving scaling laws for DP language models builds upon the framework established in~\citep{mckenna2025scaling}, while introducing three key modifications to enhance the modeling of the optimal learning rate, the estimation of loss values, and the final scaling law formulation. This approach allows for a more granular and robust understanding of the interplay between model size, training iterations, and the noise-to-batch ratio under DP constraints.

\paragraph{Explicit Modeling of the Optimal Learning Rate.}

A primary departure from previous work is the explicit modeling of the optimal learning rate. Instead of treating the learning rate as a hyperparameter to be optimized through a grid search for each configuration, we model its optimal value as a function of the training setup. For each experimental configuration, defined by a specific model size and noise-to-batch ratio, we conduct seven training runs, each with a different learning rate.

The resulting final loss values from these seven runs are then used to fit a quadratic function. The vertex of this parabola provides an estimate of the optimal learning rate for that configuration. This approach is based on the observation that the relationship between the learning rate and the final loss is often convex. The final scaling laws are subsequently derived using the loss values obtained from training each configuration at its modeled optimal learning rate. This two-step process allows for a more precise determination of the best achievable performance for each configuration, reducing the risk of suboptimal learning rate selection influencing the final scaling law.

\paragraph{Parametric Extrapolation of Loss Values.}

To efficiently estimate the loss across a continuous range of training iterations without relying on intermediate loss values, we adopt a parametric fitting approach. For each model configuration, we execute five separate training runs, each with a different, predetermined number of training iterations. The final loss from each of these five runs is then recorded.

We use a simple parametric form inspired by~\citep{hoffmann2022training}, namely $ L = E + \frac{A}{T^\alpha}$ where $L$ is the loss, $T$ is the number of training iterations and $E$, $A$, and $\alpha$ are the fitted parameters. We fit this function using \texttt{scipy.optimize.curve\_fit} to these five data points. This function allows for both interpolation and extrapolation of the loss to any number of iterations within a reasonable range. We find that this reduces overestimating the loss when training iterations are smaller than the experimental configuration.

\paragraph{Semi-Parametric Fit.}

Our final scaling law is constructed through a two-stage fitting process, which separately models the loss as a function of model size and training iterations as a parametric function described above.

We utilize the parametric extrapolation to generate a dense grid of loss values across a wide range of model sizes, training iterations, and noise-batch ratios. These generated data points are then used to fit separate non-parametric model following the methodology in~\citep{mckenna2025scaling} to predict the loss value for any value of model size, training iterations, and noise-batch ratios.

\section{Training Configuration}

\subsection{Compute Infrastructure}

We train on a $2 \times 16 \times 16$  configuration of TPUv6e totaling 2048 chips, with 2048-way data replication and 1-way model sharding. Using a vectorized implementation of per-example-clipping, we are able to process four examples per core in parallel and aggregate gradients computed across 64 independent iterations to produce each model update. As in other Gemma variants, we use the GSPMD partitioner for training step computation and the MegaScale XLA compiler.

\subsection{Batch Size and Number of Iterations}
Based on our scaling law and compute budget, we find a variety of configurations produce comparable loss similar to \citep{mckenna2025scaling}. Our final training configuration is in Table~\ref{tab:model_config}. After training, we find that our scaling law prediction for loss was within 1\% of the true value achieved.

\begin{table}[h]
\centering
\caption{Overview of the training hyperparameters.}
\label{tab:model_config}
\begin{tabular}{lc}
\hline
\textbf{Parameters} & \textbf{1B} \\
\hline
\textbf{Iterations} & 100,000 \\
\textbf{Expected Batch Size} & 517,989 \\
\hline
\textbf{Noise Multiplier} & 0.6143481 \\
$\epsilon$ & $2.0$ \\
$\delta$ & $1.1e^{-10}$ \\ 
\hline
\end{tabular}
\end{table}

\section{Empirical Privacy and Memorization}
One major benefit of DP is that it provably bounds the impact of an example on the trained model. In this section, we empirically assess these benefits on training-example memorization rates~\citep{carlini22quantifying,nasr2023scalable,gemmateam2024gemma,geminiteam2024gemini}. This “memorization rate”\footnote{We do not state or imply [here] that a model ``contains''
its training data in the sense that there is a copy of that data
in the model. Rather, a model memorizes attributes of its
training data such that in certain cases it is statistically able
to generate such training data when following rules and
using information about features of its training data that it
does contain.}
is defined as the ratio of generations from the model
that match its training data compared to all model
generations using the following setup. Thus, this rate assesses a model's likelihood of producing near-copies of text used in training~\citep{carlini2021extracting,biderman2023emergent,ippolito2022preventing}.

 We follow the methodology described in \citep{team2025gemma}. We subsample roughly 1M training data samples distributed uniformly across different corpora and test for discoverable extraction~\citep{nasr2023scalable} of this
content using a prefix of length 50 and a suffix of
length 50. If all tokens in the continuation match the source suffix, we denote the text as  “exactly memorized”. If the continuations matches up to an edit distance of 10\%, we denote this as “approximately memorized”.

\begin{figure}
    \centering
    \includegraphics[width=\linewidth]{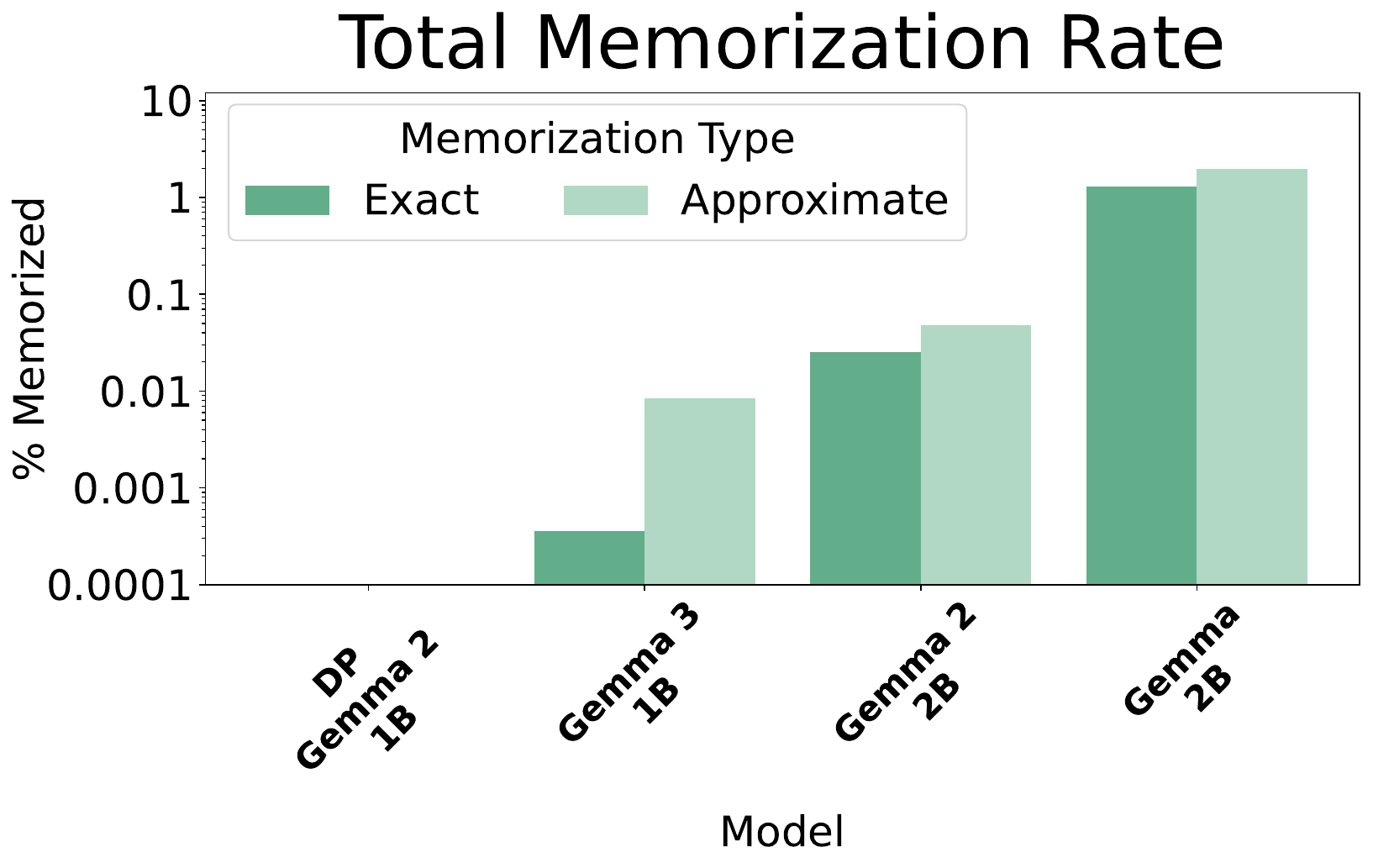}
    \caption{Total memorization rates for both exact and approximate memorization. No memorization was detected for DP Gemma.}
    \label{fig:memorization}
\end{figure}

Figure~\ref{fig:memorization} compares the memorization rates
across Gemma models ordered in reverse chronological order starting from the left. All non-DP versions of Gemma had detectable levels of memorization, that are decreasing in time due to changes in the training architecture and recipe. We were not able to detect any memorization from DP Gemma, despite using the older training architecture and recipe from Gemma 2.

\section{Conclusion}
In this work, we introduced VaultGemma, the largest open-weight language model trained from its inception with a formal DP guarantee. The development of this model was guided by our formulation of novel scaling laws for DP training, which provide a principled and quantitative framework for navigating the inherent trade-offs between model utility, privacy, and computational cost. We are releasing the model weights and our training methodology to facilitate reproducibility and encourage further research in privacy-preserving machine learning.

While our results are promising, a utility gap persists between privately and non-privately trained models. The scaling laws we present offer a clear roadmap for future research aimed at improving the performance of private models. Future work could focus on developing more efficient algorithms, exploring novel data curation strategies, and applying these principles to even larger models. VaultGemma serves as a powerful baseline and a key step toward making large-scale, provably private AI a practical reality.

\clearpage

\section{Contributions and Acknowledgments}
\textbf{Core Contributors} \\
Amer Sinha\\
Thomas Mesnard\\
Ryan McKenna\\
Daogao Liu\\
Christopher A. Choquette-Choo\\
Yangsibo Huang\\
Da Yu\\
George Kaissis\\
Zachary Charles\\
Ruibo Liu\\ 
Lynn Chua\\
Pritish Kamath\\
Pasin Manurangsi\\
Steve He\\\\
\textbf{Tech Leads}\\
Chiyuan Zhang\\
Badih Ghazi\\
Borja De Balle Pigem\\\\
\textbf{Product}\\
Prem Eruvbetine\\
Tris Warkentin\\\\
\textbf{Leads}\\
Armand Joulin\\
Ravi Kumar\\\\
\textbf{Acknowledgements}\\
Peter Kairouz\\
Brendan McMahan\\
Dan Ramage\\ 
Andreas Terzis\\

\bibliography{main}

\end{document}